\documentstyle[preprint,oldlfont,aps,doublespace,epsf]{revtex}
\begin{document}
\draft 
\title{Comment on ``A Criterion that determines foldability of proteins'' by D.Klimov and D.Thirumalai}  
\maketitle
The paper by Klimov and Thirumalai (KT) [1] presents results for protein folding 
obtained from Monte Carlo lattice model simulations 
of the type introduced  in [2].  They draw two conclusions 
from their study of 15-mers  and 27-mers.  
The first conclusion is 
that there exists a strong 
correlation between the 
folding rate of a sequence and the 
parameter $\sigma = T_f/T_{\Theta} - 1$, 
where $T_f$ is the temperature at 
which the ``chain undergoes the transition 
to the folded state'' and  $T_{\Theta}$ is a temperature, 
above 
which chain "is in an extended random coil state" 
[1].  The second conclusion is that the energy gap 
criterion for folding found by Sali et al. 
(SSK) in their 27-mer simulations [3] is 
not applicable to the systems studied by KT.  In 
this comment, we point out that the 
folding criterion used by KT is essentially the same as 
one introduced earlier [3,4] and 
we make clear that application of the energy gap criterion to 
the KT simulations is inappropriate; 
i.e., their use of the criterion is based on a 
misunderstanding of the SSK analysis.  

It was shown in [4] by a study of 13 different sequences designed to 
fold into the same structure of a cubic lattice 36-mer that at high temperatures the folding 
rate is correlated with $T_f$  (see Fig.8a of [4]), 
which in turn is related to the ``stability gap'' 
(energy difference between
the native state and the bulk of non-native, structurally dissimilar
conformations).
 Sequences of different stabilities were studied in [4]
while KT studied only well optimized sequences of 27-mers; 
hence their Fig. 2b corresponds to the 
right most part of Fig. 8a of [4]. 
The only difference between the $T_f$ 
criterion used in [4] and the $\sigma$-criterion of  KT is 
normalization by $T_{\Theta}$.  However, such  a normalization
is unlikely to improve the correlation over 
that with $T_f$ alone and it introduces
a potential source of ambiguity.
For the short polymers studied by KT  the
collapse transition is very broad so that
$T_{\Theta}$ is not well-defined [5]. 
This can be seen clearly in 
Figs. 2c and 3c of [6] where the thermodynamics of the 
collapse 
and folding transitions in short (N=16) chains 
was studied by exhaustive enumeration. 
The resulting uncertainty in $T_f/T_{\Theta}$
 makes that
in $\sigma$ greater than the 
whole range over which it varies in the KT study.

KT state that there is "no useful correlation 
between the folding time and energy gap..." in their simulations.  In this 
regard, they cite a paper [7] which was concerned with the relation between 
energy landscape and the folding mechanism of protein-like lattice chains.   The energy gap 
criterion was the subject of an 
earlier paper [3], which 
presented a detailed study of the 
folding propensities of 200 sequences.  
It was shown that the
fast-folding and 
slow-folding sequences 
in this model 
can be 
distinguished by the value of
the energy gap $\Delta_{CS}$ 
between the ground and first 
excited state for the {\em fully compact}  
($3 \times 3 \times 3$ cube) ensemble.  
The advantage of the fully compact ensemble
is that it is enumerable for 27-mers
on the cubic lattice, and can be 
used as a representative of the whole conformational 
space (see Appendix II of ref. 3). 
Also, it makes it possible to
study the relation between the
density of states and folding kinetics by 
independent calculations, i.e.
the former from exhaustive enumeration and the
latter from MC simulations. 
Finally and most important for 
the present considerations,
use of the fully compact  ensemble for analysis
eliminates conformations 
differing from the ground state by the rearrangement 
of only a few monomers so that $\Delta_{CS}$ 
characterizes the ``stability gap'' (see above).  
Further, Fig. 7 of [3] shows a correlation 
between thermodynamic 
stability, as determined from 
exhaustive enumeration of compact 
conformation, and fast 
folding. This is closely related to the 
correlation between $T_f$ and 
fast folding considered above. 
KT do not find any correlation 
between fast folding and the ``energy gap'' $\Delta$ 
in their 
simulations because their 
non-compact structures have conformations that differ from the 
native state by the position of 
only one (or a few) monomers. 
The difference between $\Delta$ and $\Delta_{CS}$, determined by SSK, 
was already  clear
from Fig.17 of [3]  (which 
is essentially duplicated by Fig. 1 of [1]). 
Thus, no relation between $\Delta$ and folding kinetics is expected.

\vspace{10pt}
\noindent
M.Karplus  and E. Shakhnovich \\
Harvard University \\ 
Department of Chemistry \\
 12 Oxford Street, Cambridge MA 02138

\vspace{5pt}

[1] D.Klimov and D.Thirumalai,
Phys.Rev.Lett {\bf 76}, 4070 (1996).

[2] E. Shakhnovich, et al, 
Phys. Rev. Lett. {\bf 67}, 1665 (1991),

[3] A.Sali, E.I.Shakhnovich, and M.Karplus,
Journ. Mol. Biol. {\bf 235}, 1614--1636, (1994).

[4] V.Abkevich, A.~Gutin, and E.Shakhnovich,
J.Chem.Phys, {\bf 101} 6052, (1994).

[5] A.Yu. Grosberg and A.R.Khohlov,
{\em Statistical Mechanics of Macromolecules}.
 AIP Press, N.Y.,N.Y, 1994.

[6] A.~Dinner, A.Sali, M.Karplus, and E.Shakhnovich,
J. Chem. Phys {\bf 101}, 1444 (1994).

[7] A.Sali, E.I.Shakhnovich, and M.Karplus,
Nature {\bf 369}, 248 (1994).

\end{document}